\documentclass[aps,preprint]{revtex4}%
\usepackage{amsfonts}
\usepackage{amsmath}
\usepackage{amssymb}
\usepackage{graphicx}%
\providecommand{\U}[1]{\protect\rule{.1in}{.1in}}

\begin{document}
\title{Cooling by heating in Quantum Optics Domain}
\author{D. Z. Rossatto$^{1}$, A. R. de Almeida$^{2,3}$, T. Werlang$^{1}$, C. J.
Villas-Boas$^{1}$, N. G. de Almeida$^{3}$}
\affiliation{$^{1}$Departamento de F\'{\i}sica, Universidade Federal de S\~{a}o Carlos,
13565-905, S\~{a}o Carlos, S\~{a}o Paulo, Brazil}
\affiliation{$^{2}$UnUCET - Universidade Estadual de Goi\'{a}s, 75132-903, An\'{a}polis
(GO), Brazil}
\affiliation{$^{3}$Instituto de F\'{\i}sica, Universidade Federal de Goi\'{a}s, 74.001-970,
Goi\^{a}nia - GO, Brazil}

\begin{abstract}
A class of Hamiltonians that are experimentally feasible in several contexts
within Quantum Optics area and which lead to the so-called cooling by heating
for fermionic as well as for bosonic systems have been analyzed numerically.
We have found a large range of parameters for which cooling by heating can be
observed either for the fermionic system alone or for the combined fermionic
and bosonic systems. Finally, analyzing the experimental requirements we
conclude that cooling by heating is achievable with nowadays technology,
especially in the context of trapped ions/cavity QED, thus contributing to
understand this interesting and counter intuitive effect.

\end{abstract}
\maketitle

Recently, two schemes were proposed for cooling by heating \cite{Mari2012,
Cleuren2012}, \textit{i.e}., a given physical system in contact with a thermal
reservoir has decreased its energy when one increases the temperature of the
reservoir. A. Mari \textit{et al}. \cite{Mari2012} introduced the idea of
cooling a quantum system using incoherent thermal light. They proposed a
scheme based on an optomechanical system, demonstrating that by driving the
system with a thermal noise the interaction with other modes can be enhanced
to assist in cooling the optomechanical system. In a another work, B. Cleuren,
B. Rutten, and C. Van den Broeck \cite{Cleuren2012} proposed a scheme to cool
a system powered by photons. Their systems are based on a nanosized solid
state device, with no moving parts and no net electric currents, which can be
refrigerated directly by using thermal photons.

In this brief report we investigate numerically a class of well known
Hamiltonians in the quantum optics domain and show that these Hamiltonians can
lead to cooling by heating\textbf{. }Differently from both schemes above,
which investigate the cooling by heating in solid state or optomechanical
devices, our work brings this striking effect to the quantum optics context
where techniques to manipulate systems at the individual atomic and bosonic
scale is daily presented, thus opening the possibility to experimentally
observe this phenomenon in a very controllable scenario.

\textit{Model. }In order to find out a system which allows us to cool it by
raising the temperature of its reservoir, firstly we must note that all the
systems which thermalize with the environment can not present this phenomenon
(for example a single two-level atom or a single bosonic mode interacting with
a thermal reservoir). Thus, to see cooling by heating some external force must
be employed to drive the system out of equilibrium with the environment. To
this end we explore the well known generalized anti-Jaynes-Cummings model
(JCM) (which will be derived bellow), in the coupling regime where the
effective Rabi frequency (atom-boson coupling) is much smaller than the
bosonic and atomic transition frequencies. To implement such Hamiltonians in
the trapped ions domain, for instance, one can use a two-level ion
characterized by the transition frequency $\omega_{0}$ between the ground
$\left\vert g\right\rangle $ and excited $\left\vert e\right\rangle $ states
and trap frequency $\nu$ (bosonic mode) \cite{Wineland03}. The transition
$\left\vert g\right\rangle $ $\leftrightarrow$ $\left\vert e\right\rangle $ is
driven by a classical field of frequency $\omega_{L}$, wave vector
$k_{L}=\omega_{L}/c$, and Rabi frequency $\Omega$ \cite{Wineland03}. In the
Schr\"{o}dinger picture, the Hamiltonian which describes such a system reads
($\hbar=1$) $H=H_{f}+H_{a}+H_{int}\left(  t\right)  $, with $H_{f}=\nu
a^{\dagger}a$, $H_{a}=\omega_{0}\sigma_{z}/2$ and
\begin{equation}
H_{int}\left(  t\right)  =\frac{\Omega}{2}\sigma_{-}e^{i\left(  k_{L}%
\widehat{x}+\omega_{L}t\right)  }+H.c.\text{,} \label{00}%
\end{equation}
$\sigma_{+}$($\sigma_{-}$) being the usual raising (lowering) Pauli operator
for a two-level atomic system, $\sigma_{z}=\sigma_{+}\sigma_{-}-\sigma
_{-}\sigma_{+}$, $a$ ($a^{\dagger}$) is the annihilation (creation) operator
in the Fock space for the bosonic mode (vibrational motion of the ion),
\ $H.c.$ means Hermitian conjugate, $k_{L}\widehat{x}=$ $\eta_{L}\left(
a+a^{\dagger}\right)  $, with $\eta_{L}=$ $k_{L}/\sqrt{2m\nu}$ being the
Lamb-Dicke parameter \cite{Wineland03}. Working in the limit $\eta_{L}\ll1$
and applying the rotating wave approximation, the Hamiltonian $H$ in the
interaction picture can be written as \cite{Wineland03}
\begin{equation}
H_{I}=g_{k}(\sigma_{-}a^{k}+\sigma_{+}a^{\dagger k}),\text{ }k=0,1,2.
\label{1}%
\end{equation}
By adjusting the frequency $\omega_{L}$ on resonance with the two-level ion we
can have the carrier interaction ($k=0$, $g_{0}=\Omega/2$); adjusting
$\delta=\omega_{L}-\omega_{0}=k\nu$, we can also have the first ($k=1$,
$g_{1}=i\Omega\eta_{L}/2$) and second ($k=2$, $g_{2}=-\Omega\eta_{L}^{2}/4$)
blue sideband interactions \cite{Wineland03}. The quantum of vibrational
energy of the center of mass of the ion is then described by $a^{\dagger}a$.
In quantum optics area, the dynamics of this model under Born and Markov
approximations (weak system-reservoir coupling) is provided by the master
equation formalism \cite{Scully}, which for the Hamiltonian (\ref{1}) reads%
\begin{align}
\frac{\partial\rho}{\partial t}  &  =-i\left[  H_{I},\rho\right]
+\kappa\left(  n_{th}+1\right)  \mathcal{D}[a]\rho+\kappa n_{th}%
\mathcal{D}[a^{\dag}]\rho\nonumber\\
&  +\gamma\left(  m_{th}+1\right)  \mathcal{D}[\sigma_{-}]\rho+\gamma
m_{th}\mathcal{D}[\sigma_{+}]\rho\label{2}%
\end{align}
where $\kappa$ and $\gamma$ are the spontaneous emission rates for the
vibrational motion and internal levels of the ion, respectively, $n_{th}$
($m_{th}$) is the mean number of phonons (photons) of the reservoir coupled to
the vibrational mode (internal levels of the ion), and $\mathcal{D}%
[A]\rho\equiv2A\rho A^{\dagger}-A^{\dagger}A\rho-\rho A^{\dagger}A$
\cite{Lindblad}.

Below we proceed to solve numerically (or analytically, for the carrier
interaction) this master equation to obtain the steady state of the system (at
$t\rightarrow\infty$ or $\partial\rho/\partial t=0$), in order to be able to
calculate the corresponding thermodynamical properties. To this aim, we
firstly have to note that the master equation can give rise to an infinity set
of coupled differential equations for the elements of the density matrix of
the whole system. Then, to solve it numerically, first we must truncate the
Fock basis of the bosonic field somewhere. This truncation depends on the mean
number of excitations in the vibrational mode, \textit{i.e.}, the matrix
elements corresponding to highly excited Fock states (compared to the mean
number of excitation of the vibrational mode) must be virtually zero. We then
integrate numerically the system of coupled differential equations for the
elements of the density matrix of the system following the method presented in
\cite{Tan99}. As we are working with two distinct reservoirs, there will be
different response functions: one for the atom and another for the vibrational
mode \cite{Zia02}. Working with the Hamiltonian (\ref{1}) we can distinguish
three situations where cooling by heating (i.e., by raising the temperature of
the reservoir) can be observed:

\textit{(i) }looking at the variation of the internal energy of the ion only;

\textit{(ii)} looking at the variation of the vibrational energy of the ion;

\textit{(iii)} looking at the variation of both internal and vibrational
energies of the ion.

\textit{Variation of the internal energy of the ion only}. Firstly we assume a
carrier interaction ($k=0$) in Eq.(\ref{1}) which corresponds to a single
two-level ion driven by a classical field. Since the dynamics of the system
does not involve the vibrational mode, we can fix $\kappa=0$ without loss of
generality. From Eq.(\ref{2}), we can easily obtain the average internal
energy $E_{a}=\left\langle H\right\rangle =\left\langle H_{a}+H_{INT}%
\right\rangle =\left\langle H_{a}\right\rangle $ of the ion in the steady
state as a function of the mean number of thermal photons (temperature of the
reservoir). Then we can calculate the response function ($C_{a}$) of the
internal energy with respect to the temperature of its reservoir ($T$) which
we define as
\begin{equation}
C_{a}=\frac{\partial E_{a}}{\partial T}. \label{RF}%
\end{equation}
This equation resembles the usual definition of specific heat. However, note
that the temperature appearing in the equation above is the one of the
reservoir, which is different of the effective temperature of the system since
it is not in thermal equilibrium with its environment. With the steady state
solution for the internal energy $E_{a}$, we can analytically derive the
response function
\begin{align}
C_{a}  &  =-2k_{B}m_{th}\left(  m_{th}+1\right) \nonumber\\
&  \times\left[  \ln\left(  \frac{m_{th}+1}{m_{th}}\right)  \right]  ^{2}%
\frac{\left[  2\left(  g_{0}/\gamma\right)  ^{2}-\left(  2m_{th}+1\right)
^{2}\right]  }{\left[  2\left(  g_{0}/\gamma\right)  ^{2}+\left(
2m_{th}+1\right)  ^{2}\right]  ^{2}}, \label{cv1}%
\end{align}
$k_{B}$ being the Boltzmann constant. Clearly, we see from Eq.(\ref{cv1}) that
$C_{a}\leq0$ for%
\begin{equation}
m_{th}\leq\frac{1}{\sqrt{2}}\frac{g_{0}}{\gamma}-\frac{1}{2}.
\label{rangenegative}%
\end{equation}
Note that $C_{a}\rightarrow0$ when $m_{th}\rightarrow0$ (similar to what
occurs for the third law of thermodynamics) or $m_{th}\rightarrow\infty$
(system saturation). For a sample of $N$ non-interacting atoms \cite{Natoms},
the response function is $C_{N}=NC_{a}$ and then, the negative response can be
observed even for an ensemble of two-level atoms.

\textit{Variation of the vibrational and internal energy of the ion}.
Considering the ion coupled to the vibrational mode, we note that using the
average energy of an individual system instead of the total energy (sum of the
subsystems and interaction average energies) does not change our conclusions,
since there is a region where the response function, Eq.(\ref{RF}), becomes
negative for both systems simultaneously and $Tr\left(  H_{I}\rho\right)  =0$
in all cases studied here, $H_{I}$ being the interaction Hamiltonian
(\ref{1}). To see that this is so, in Fig. 1(a) ($k=1$) and Fig. 2(a) ($k=2$)
we plot the stationary average energy for both the bosonic mode ($\left\langle
H_{f}\right\rangle /\nu$) and the atomic system ($\left\langle H_{a}%
\right\rangle /\omega_{0}$). In all simulations we assumed zero atomic energy
in the lower state $\left\vert g\right\rangle $. To reliably calculate the
region where cooling by heating can occur, we have limited our numerical
analysis to the range $0\leq g_{k},\kappa\leq2\gamma$ ($k=1,2$). Also, in all
figures, the average atomic energy was multiplied by a factor of ten for the
sake of clarity. Assuming the two reservoirs at a common temperature
($m_{th}=n_{th}$), first we set $\kappa=0.1\gamma$ and $g_{1}=1.0\gamma$ in
Fig. 1(a), and $\kappa=0.1\gamma$ and $g_{2}=0.2\gamma$ in Fig. 2(a).
Remarkably, note from those figures that there is a region where the response
to the rising reservoir temperature of both atomic and bosonic system is
negative (falling energy), thus supporting our assertion that cooling by
heating can be observed even if we adopt a definition, different from
Eq.(\ref{RF}), taking into account the total energy. Also, note that the final
temperature of the bosonic system differs from that of its reservoir $\left(
\left\langle a^{\dag}a\right\rangle \neq n_{th}\right)  $, thus indicating the
existence of non-equilibrium steady states \cite{Lynden-Bell99,Zia02}. It is
important to mention that it is not surprisingly to have a
\textit{non-equilibriun }steady state once the system is driven by an external
force (the external laser).

In Fig. 1(b) (Fig. 2(b)) we plot the response function versus $n_{th}$ to the
bosonic and the atomic systems for the model $k=1$ ($k=2$). Both figures show
that cooling by heating for the bosonic system can occur in a wider region
than that for the atomic system.

Let us now explore the fact that the reservoirs for the atomic and bosonic
systems can have different mean number of thermal photons. This is
particularly relevant for trapped ion experiments since the transition
frequency $\omega_{0}$ (of the order of few GHz) of the electronic levels
involved are usually much bigger than the frequency $\nu$ of the ionic motion
($\sim$MHz) \cite{Meekhof96}, resulting in different mean number of thermal
photons for the electronic levels and ionic motion for a given temperature. In
Fig. 3(a) we show the behavior of the response function for the atomic system
when the mean photon number of the bosonic reservoir is fixed at $n_{th}=0.0$,
$1.0$, $2.0$, for the model $k=1$, using the same parameters as those in Fig.
1. Note that cooling by heating can occur when $m_{th}\lesssim1$ irrespective
of the fixed $\ n_{th}$. Fig. 3(b) does the same for the model $k=2$, with the
parameters used in Fig. 2. Cooling by heating can now occur when
$m_{th}\lesssim0.5$.

It is noteworthy that when we fix the average number of thermal photons for
the atomic reservoir and investigate the behavior of the response function to
the bosonic system as a function of temperature, we do not see regions where
cooling by heating can occur. Besides, for the range of parameters used in our
numerical simulations, we have found regions where cooling by heating can
occur in the atomic system for some values of the effective Rabi frequency
$g_{k}$, irrespective of the ratio $\kappa/\gamma$.

On the other hand, to the system under study, to observe cooling by heating
for the bosonic system for some effective Rabi frequency $g_{k}$, not only the
reservoirs must have the same average photon number ($m_{th}=n_{th}$) but the
atomic decay must be stronger than the bosonic mode decay, which our numerical
simulations point to the ratio $\kappa/\gamma\lesssim0.3$ for the model $k=1$
and $\kappa/\gamma\lesssim0.4$ for the model $k=2$. In a first moment, one
could think that we should have the same response regardless the bosonic or
the fermionic system, once the equation of motion (3) is completely symmetric
on the fermionic and on the bosonic operators. However, the nature of those
operators are completely different, i.e., the fermionic operators are
restricted to a two-dimension Hilbert space while the bosonic ones are in a
infinite Hilbert space. So, the physical difference is the number of
accessible states of each subsystem: the fermionic subsystem has only two
accessible states and the bosonic subsystem can access infinite states.

\textit{Experimental proposal}: We now comment on the parameters appearing in
the effective Hamiltonians discussed above and how cooling by heating could be
observed with the nowadays technology. In the trapped ions domain, for
instance, Hamiltonians Eq.(\ref{1}) were obtained and used to engineer
nonclassical motional states \cite{Wineland03}. For the anti-JCM ($k=1$) and
the so-called two-phonon anti-JCM ($k=2$), the effective couplings are,
respectively, $\left\vert g_{1}\right\vert =\left\vert \eta_{L}\Omega
\right\vert /2$ and $\left\vert g_{2}\right\vert =\left\vert \eta_{L}%
^{2}\Omega\right\vert /4$, where $\eta_{L}$ is the Lamb-Dicke parameter and
$\Omega$ is the Rabi frequency of the classical field driving the two-level
ion, which can easily be adjusted \cite{Wineland03}. For the hyperfine ground
states of a single $^{9}Be^{+}$ ion, one can adjust $\eta_{L}$ $=$ $0.2$, thus
lying in the Lamb-Dicke regime \cite{Meekhof96}. We note that typical starting
values of the average number of thermal phonons in the mode of interest are
between $0$ and $2$ and the decay rate of the vibrational motion of the ion
can be much smaller than $g_{k}$ \cite{Turchette00}. Thus, a trapped ion seems
to be an appropriate physical system to observe cooling by heating. The
existence of cooling by heating in this context provides an interesting and
counter-intuitive application: the ion motion can be reduced as the reservoir
temperature increases. The anti-JCM can also be engineered in a cavity QED
setup \cite{marcelo}. As the usual atom-field coupling in the microwave domain
is $\lambda\sim10^{5}s^{-1}$, the effective coupling for the anti-JCM can be
$g_{1}\sim10^{3}s^{-1}$ \cite{marcelo}. The cavity decay rate $\kappa$ ranges
from $10s^{-1}$ to $10^{2}s^{-1}$ \cite{Gleyzes07} and, therefore, we easily
attain the condition $0\leq g_{k}/\kappa\lesssim10$. Taking into account
realistic temperatures, the effective mean occupation number at the microwave
frequency has to be $n_{th}\sim0.7$, according to QED cavity experiments
\cite{harocheRev2001}. This mean number of thermal photons can be reduced down
to $0.1$ by sending atoms resonantly with the cavity mode to absorb the
thermal field \cite{harocheRev2001}.

\textit{Conclusion}. We studied a class of Hamiltonians well-known in the
quantum optics domain and showed that cooling by heating can occur for a large
range of parameters, including some achievable by present day techniques. We
numerically solve the master equation and calculate the response function of
the internal energy for systems interacting with a thermal bath when varying
their corresponding reservoir temperature: $\ a)$ a single two-level atom (or
even a sample of $N$ two-level atoms) driven by a classical field and $b)$ a
bosonic mode interacting with a two-level ion/atom. We hope this work will
trigger a search for experimental verification of cooling by heating in the
quantum optics area, thus strongly contributing to understand this interesting
and counter intuitive effect.

\begin{acknowledgments}
The authors acknowledge the financial support from the Brazilian agencies CNPq
and Brazilian National Institute of Science and Technology for Quantum
Information (INCT-IQ). C. J. V. B. also acknowledges support from FAPESP
(Proc. 2012/00176-9).
\end{acknowledgments}

\bigskip

\textbf{Figure Caption}

Fig. 1: (color online) (a) Average energy for atomic ($\left\langle
H_{a}\right\rangle /\omega_{0}$ - dashed line (x10)) and bosonic systems
($\left\langle H_{f}\right\rangle /\nu$ - solid line) versus a common mean
number of thermal photons $n_{th}=m_{th}$, for the model $k=1$. (b) Response
function versus mean\ number of thermal photon. The cooling by heating can
occur for $m_{th}=n_{th}\lesssim1.4$ to the bosonic and $m_{th}=n_{th}%
\lesssim0.9$ to the atomic system. The parameters used are $\kappa=0.1\gamma$
and $g_{1}=1.0\gamma$.

\bigskip

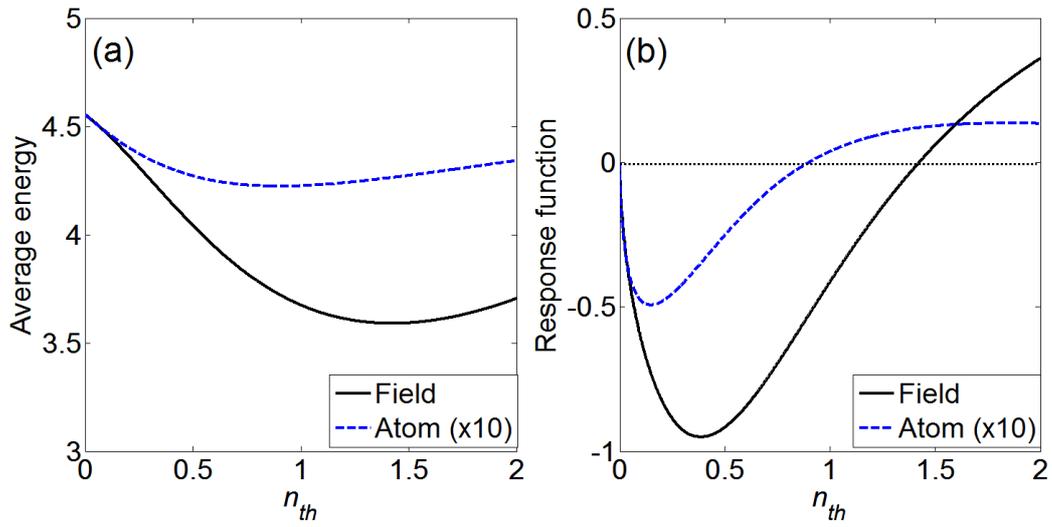
Fig. 2: (color online) (a) Average energy for atomic (dashed line (x10)) and
bosonic systems (solid line) versus a common mean number of thermal photons
$n_{th}=m_{th}$, for the model $k=2$. The mean value of the interaction
Hamiltonian $H_{I}$ (not shown in this figure) is always zero. (b) Response
function versus mean number of thermal photons. The cooling by heating can
occur for $m_{th}=n_{th}\lesssim1.2$ to the bosonic and $m_{th}=n_{th}%
\lesssim0.4$ to the atomic system. The parameters used are $\kappa=0.1\gamma$
and $g_{2}=0.2\gamma$.

\bigskip

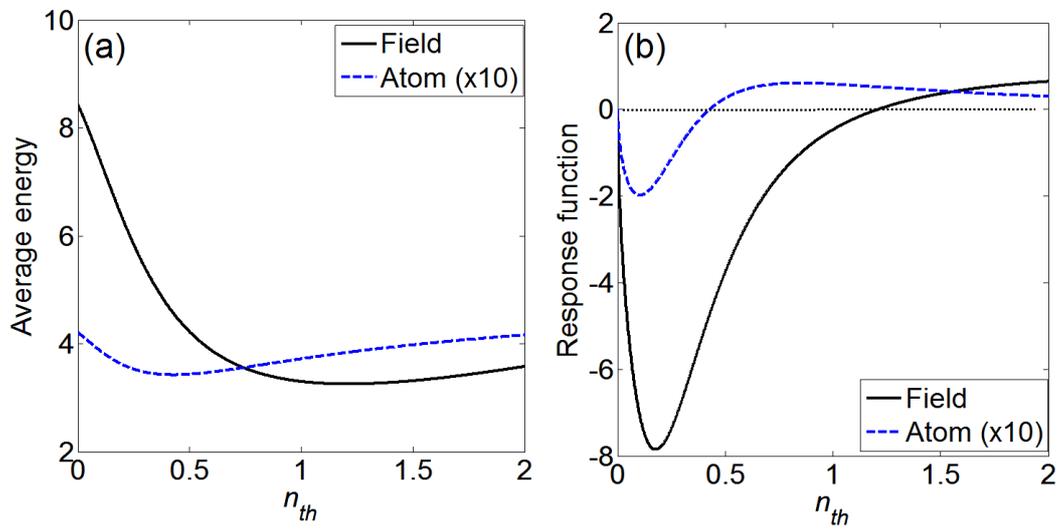
Fig. 3: (color online) Response function to the atomic system versus the
average photon number $m_{th}$ of its reservoir when fixing the average photon
number of the bosonic reservoir: $n_{th}=0.0$ (solid line), $n_{th}=1.0$
(dashed line) and $n_{th}=2.0$ (dotted line), for the model (a) $k=1$, using
$\kappa=0.1\gamma$ and $g_{1}=1.0\gamma$; and (b) $k=2$, using $\kappa
=0.1\gamma$ and $g_{2}=0.2\gamma$.

\bigskip

\begin{thebibliography}{99}                                                                                               %


\bibitem {Mari2012}A. Mari and J. Eisert, Phys. Rev. Lett. \textbf{108},
120602 (2012).

\bibitem {Cleuren2012}B. Cleuren, B. Rutten, and C. Van den Broeck, Phys. Rev.
Lett. \textbf{108}, 120603 (2012).

\bibitem {Wineland03}D. Leibfried \textit{et al.,} Rev. Mod. Phys.
\textbf{75}, 281 (2003).

\bibitem {Scully}H.-P. Breuer and F. Petruccione, \textit{The theory of Open
Quantum Systems} (Oxford Unversity Press, Oxford, 2007).

\bibitem {Lindblad}G. Lindblad, Commun. Math. Phys. \textbf{48}, 119 (1976)

\bibitem {Tan99}S. M. Tan, J. Opt. B: Quantum Semiclass. Opt. \textbf{1}, 424 (1999).

\bibitem {Zia02}R. K. P. Zia, E. L. Praestgaard, and O. G. Mouritsen, American
Journal of Physics \textbf{70,} 384 (2002).

\bibitem {Natoms}The Hamiltonian for $N$ identical non-interacting atoms is
obtained by replacing $\sigma_{+}\rightarrow S_{+}=\sum_{i=1}^{N}\sigma
_{+}^{i}$ in Eq. (\ref{1}) for $k=0$ $\left(  S_{-}=S_{+}^{\dagger}\right)  $.
Here $\sigma_{+}^{i}$ is the Pauli operator acting on the j-th atom. The
calculation of the response function for $N$ atoms is almost identical to the
calculation performed for a single atom

\bibitem {Lynden-Bell99}D. Lynden-Bell, Physica A \textbf{263}, 293 (1999).

\bibitem {Meekhof96}D. M. Meekhof \textit{et al.,} Phys. Rev. Lett.
\textbf{76}, 1796 (1996).

\bibitem {Turchette00}Q. A. Turchette \textit{et al}., Phys. Rev. A
\textbf{61}, 063418 (2000).

\bibitem {marcelo}M. Fran\c{c}a Santos, E. Solano, and R. L. de Matos Filho,
Phys. Rev. Lett. \textbf{87,} 093601 (2001).

\bibitem {Gleyzes07}S. Gleyzes \textit{et al.,} Nature \textbf{446}, 297 (2007).

\bibitem {harocheRev2001}J. M. Raimond, M. Brune, and S. Haroche, Rev. Mod.
Phys. \textbf{73}, 565 (2001)..
\end{thebibliography}
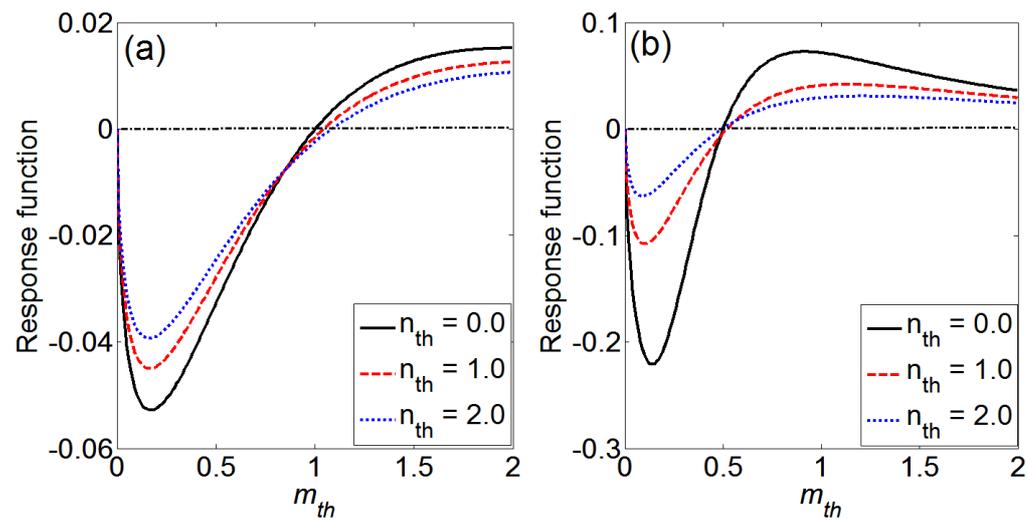
\end{document}